\documentclass[conference]{IEEEtran}
\IEEEoverridecommandlockouts
\usepackage{cite}
\usepackage{amsmath,amssymb,amsfonts}
\usepackage{algorithmic}  
\usepackage[algo2e]{algorithm2e} 
\usepackage{setspace}

\usepackage{graphicx}
\usepackage{textcomp}
\usepackage{xcolor}
\usepackage{units}
\usepackage{multirow}
\usepackage{colortbl}
\usepackage{soul}
\usepackage{ragged2e}
\usepackage[left=0.625in, right=0.625in,top=0.75in,bottom=1in]{geometry}
\usepackage{subfig}

\usepackage{amsthm}
\usepackage{thmtools, thm-restate}
\usepackage{algorithm}

\usepackage{hyperref}

\DeclareMathOperator*{\argmax}{arg\,max}

\def\BibTeX{{\rm B\kern-.05em{\sc i\kern-.025em b}\kern-.08em
    T\kern-.1667em\lower.7ex\hbox{E}\kern-.125emX}}
\begin{document}

\title{Semantic Forwarding for Next Generation Relay Networks}

\author{Enes Arda, Emrecan Kutay and Aylin Yener
\\ INSPIRE@OhioState Research Center 
\\Dept. of Electrical and Computer Engineering
\\ The Ohio State University
\\ arda.2@osu.edu, kutay.5@osu.edu, yener@ece.osu.edu 
}

\newgeometry{left=0.625in, right=0.625in,top=0.75in,bottom=1in}

\maketitle

\begin{abstract} We consider cooperative semantic text communications facilitated by a relay node. We propose two types of semantic forwarding: semantic lossy forwarding (SLF) and semantic predict-and-forward (SPF). Both are machine learning aided approaches, and, in particular, utilize attention mechanisms at the relay to establish a dynamic semantic state, updated upon receiving a new source signal. In the SLF model, the semantic state is used to decode the received source signal; whereas in the SPF model, it is used to predict the next source signal, enabling proactive forwarding. Our proposed forwarding schemes do not need any channel state information and exhibit consistent performance regardless of the relay's position. Our results demonstrate that the proposed semantic forwarding techniques outperform conventional semantic-agnostic baselines. 
\end{abstract}  


\section{Introduction}
\label{sec:intro}
Exact reconstruction of messages represented by bit sequences at the receiver, after having been transmitted in a noisy channel, has been the primary objective of communication systems \cite{Shannon}. This semantic-agnostic approach has been the foundation of today's digital communication systems owing to its simplicity. Future applications however, require more efficient utilization of resources to deliver highly reliable and low-latency decisions. These applications can benefit from content-aware approaches to communications, of which semantic communications is increasingly becoming prominent for 6G \cite{sem_com_review_1, emrecan_semantic_overview}. In a broad sense, semantic communications aims to convey the meaning to the receiver, rather than requiring an exact reconstruction of what is sent \cite{guler2014semantic, semcom_game}. As such, it is especially useful for communication scenarios where the end node has a task to accomplish \cite{beyond_bits_overview}. 

Semantic communications leveraging deep learning (DL) architectures has recently been explored extensively\cite{zhijin_deepsc, sagduyu_task_oriented_nextg_23, sagduyu_is_semantic_communication_secure_23}.  Utilization of pre-trained language models for codebook construction in text compression has been investigated in \cite{emrecan_icc_ws}, and shown to exhibit better time efficiency in classification over point-to-point wireless channels\cite{emrecan_icassp}. Transformer architectures have been employed in reference \cite{zhijin_deepsc} for point-to-point text communications to extract semantic information from source messages and reconstruct a message with a similar meaning at the destination. Image communication in multiple access channels has been studied in  \cite{gunduz_semantic_mac}, where edge devices, equipped with convolutional neural networks (NNs), collectively engage in image retrieval by transmitting their images to the edge server. In interference channels, reference \cite{zhijin_multiuser_task} considers communication of both image and text, where each transmitter-receiver pair communicates independently over a shared medium for tasks such as machine translation and image retrieval. While these and other recent studies consider different communication models, the use of semantic communications in cooperative networks remains largely unexplored, which we consider in this paper.

Cooperative communications has been shown to boost the performance by means of nodes transmitting messages for one another\cite{erkip_cooperation_part_1}. A building block for cooperative communications is a three node network where a relay assists in communications between a sender and a receiver\cite{cover_elgamal_79}. In particular a relay can decode, compress or amplify the source signal it receives and forward to the destination with the goal of helping boost reliable communication rate, i.e., exact reconstruction of the source signal \cite{relay_review}. 


In this paper, we consider cooperative semantic communications in a three-node relay network. Focusing on text communications, we propose two machine-learning aided techniques for semantic forwarding at the relay. Both leverage attention-based transformer architectures to integrate semantics into communication. The key contributions of this paper are the two semantic forwarding techniques we introduce: semantic lossy forwarding (SLF) and semantic predict-and-forward (SPF). Both form a semantic state at the relay node based on the previously received signals through an attention mechanism. In the SLF approach, the relay node uses this state to decode, re-encode, and forward the received source signal. The state is then updated for the upcoming signals. Conversely, the SPF model leverages this state not for decoding the received signal but for predicting the next source signal. Our learning-based approach does not need any channel state information and exhibits consistent performance for varying relay location for a fixed source-destination distance. The simulation results over AWGN and Rayleigh fading channels demonstrate improved syntactic and semantic fidelity achieved by the proposed models, as measured by BLEU and semantic similarity scores. 

\section{System Model}
\label{sec:system_model}
\noindent {\it Notation:} In the remainder of the paper, we will refer to operations and signals as  $(.)_A^t$ where $A \in \{S, R, D\}$ denoting $\{\text{Source}, \text{Relay}, \text{Destination}\}$ and $t$ is the time index.

\begin{figure}
\centering
\includegraphics[width=2.5in]{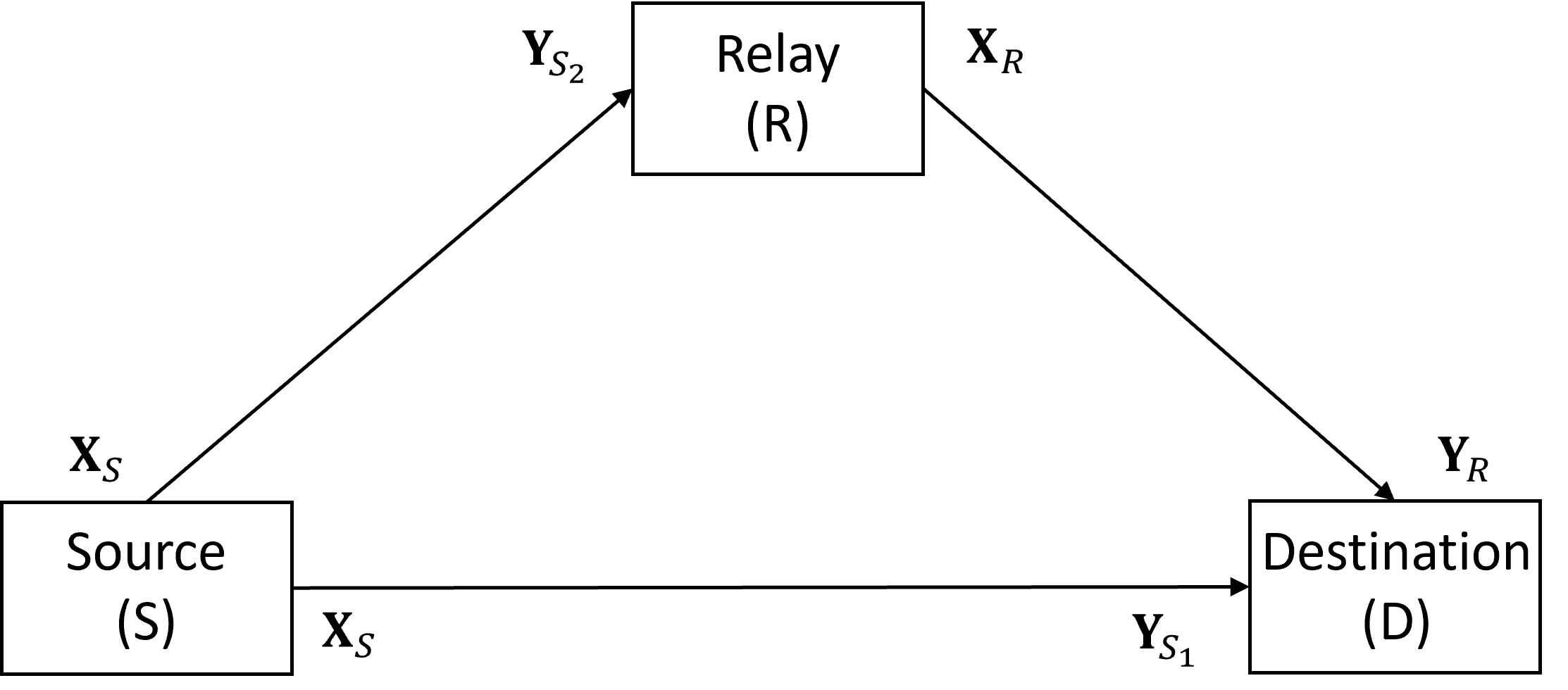} 
\caption{A three-node relay network.}
\label{fig:problem_formulation} 
\end{figure}

We consider a three-node network with orthogonal components, which consists of a source node, a relay node, and a destination node, as illustrated in figure \ref{fig:problem_formulation}. The primary objective is to transmit text messages from the source to the destination by leveraging an intermediate relay. Specifically, we assume that text messages consisting of a maximum of $L$ tokens are to be transmitted, denoted as $\mathbf{s} = \{w^1, w^2, ..., w^T\}$, where $T \leq L$. We investigate this setup over additive white Gaussian noise (AWGN) and Rayleigh fading channels. For the channel input $\mathbf{X} \in \mathbb{C}^{\mathcal{D}}$, we have the following channel output $\mathbf{Y} \in \mathbb{C}^{\mathcal{D}}$.


\begin{equation}
    \label{eq:channel_output}
    \mathbf{Y} = h\mathbf{X} + \mathbf{Z}
\end{equation}

In equation \ref{eq:channel_output}, $\mathbf{Z}$ is the additive noise component with $\mathbf{Z} \sim \mathbb{C}\mathcal{N}(\mathbf{0}, {\sigma_Z}^2\mathbf{I}_{\mathcal{D} \times \mathcal{D}})$, where ${\sigma_Z}^2 = N_0 W$. 
 $h$ represents channel gain consisting of Rayleigh fading and signal attenuation. We simulate attenuation with a 
path loss exponent $\alpha$ and the distance of the link $d_{\text{Link}}$. 
The relay is located on a line between the source and destination. 

\section{Semantic Relaying}
\label{sec:semantic_relay_model}
\begin{figure*} 
  \includegraphics[width=\textwidth]{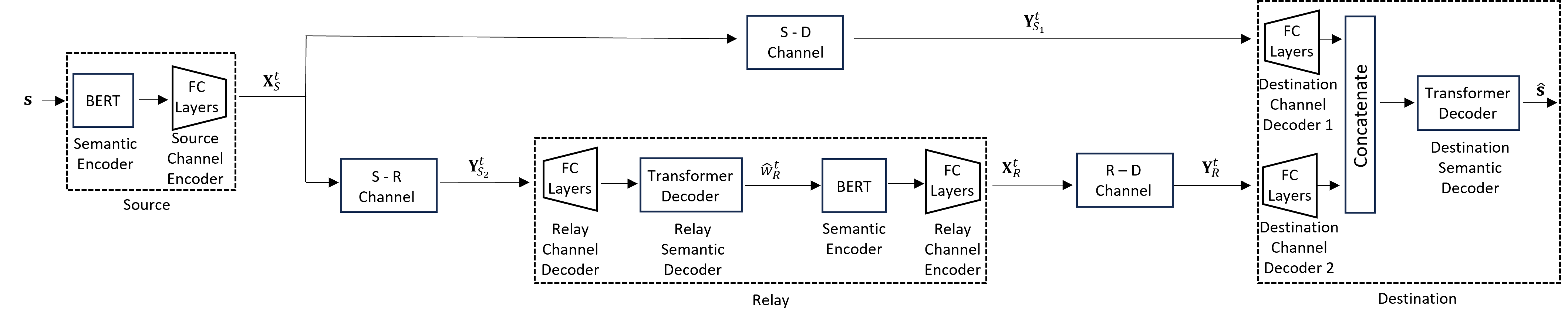}
  \caption{Block diagram of proposed single-relay cooperative semantic models.}
  \label{fig:system_model}
\end{figure*}

In this section, we present two new forwarding techniques: semantic lossy forwarding (SLF) and semantic predict-and-forward (SPF).
Both techniques share the same block diagram, as illustrated in figure \ref{fig:system_model}, but differ in their processing at the relay, as explained in sections \ref{sec:semantic_lossy_forwarding} and \ref{sec:semantic_predict_and_forward}.

\subsection{Semantic Lossy Forwarding (SLF)}
\label{sec:semantic_lossy_forwarding}
 At the source node, we extract semantic information of the entire message using a pre-trained BERT model, whose operation is denoted by $S(.)$. This information is transmitted on a token-by-token basis through the channel encoder. This process can be described as follows.

\begin{equation}
    \label{eq:source_process}
    \mathbf{X}_S^t = \Gamma_{S} \biggl(S^t \bigl(\mathbf{s} \bigl) \biggl)
\end{equation}

In equation \ref{eq:source_process}, $\Gamma_{S}(.)$ represents the channel encoding operation at the source node, and $S^t \bigl(\mathbf{s} \bigl)$ is the embedding of the token at index $t$. The encoded signal is then broadcasted to the relay and destination nodes through the channel described in equation \ref{eq:channel_output}. At the relay node, the received signal, denoted as $\mathbf{Y}_{S_2}^t$ in figure \ref{fig:system_model}, is decoded as follows. 

\begin{equation}
    \label{eq:slf_relay_decoding}
    \hat{w}_{R}^t = S_{R}^{-1} \biggl(\Gamma_{R}^{-1} \Bigl(\mathbf{Y}_{S_2}^t \Bigl), \mathbb{S}_{R}^t \biggl)
\end{equation}

In equation \ref{eq:slf_relay_decoding}, $\Gamma_{R}^{-1}(.)$ and $S_{R}^{-1}(.)$ represent the channel and semantic decoding operations at the relay, respectively. For semantic decoding, the output of channel decoder is used in conjunction with the relay's semantic state, denoted as $\mathbb{S}_{R}^t$. This semantic state is comprised of previously decoded tokens up to index $t$ and creates a context through the attention mechanism to semantically decode the received signal. This means that each decoded token at the relay impacts the future decoding operations, and the state is continually updated as new source signals are received. The decoded token $\hat{w}_{R}^t$ is then re-encoded following a procedure similar to that used at the source. We have 

\begin{equation}
    \label{eq:relay_encoding}
    \mathbf{X}_R^t = \Gamma_{R}\biggl(S^t\Bigl(\mathbf{\hat{s}}_{R}^t\Bigl)\biggl)
\end{equation}
where $\mathbf{\hat{s}}_{R}^t = \{\hat{w}_{R}^1, \hat{w}_{R}^2, \ldots, \hat{w}_{R}^t\}$. In equation \ref{eq:relay_encoding}, semantic encoding of $\hat{w}_{R}^t$ is performed with limited knowledge compared to the source node, despite employing the same BERT model as in equation \ref{eq:source_process}. At the source node, the BERT model has access to the entire message. By contrast, at the relay node, only causally decoded source signals, represented as tokens denoted by $\mathbf{\hat{s}}_{R}^t$ in equation \ref{eq:relay_encoding}, are available. That is, the decoded tokens can be noisy unlike the one at the source node, i.e., $\mathbf{\hat{s}}_{R}^t \neq \{w^1, w^2, \ldots, w^t\}$.

At the destination node, received signals from the source and the relay are decoded by separate channel decoders. The outputs of these channel decoders are concatenated and then provided as input to the semantic decoder for reconstructing the source token $w^t$ as $\hat{w}^t$.

\begin{equation}
    \label{eq:destination_decoding}
    \hat{w}^t = S_{D}^{-1} \biggl(\Gamma_{D}^{-1} \Bigl(\mathbf{Y}_{S_1}^t, \mathbf{Y}_R^t \Bigl), \mathbb{S}_{D}^t \biggl)
\end{equation}

In equation \ref{eq:destination_decoding}, $\Gamma_{D}^{-1}(.)$ and $S_{D}^{-1}(.)$ denote the channel decoding and semantic decoding operations at the destination node, respectively. Similar to the decoding process at the relay node, decoding at the destination node also incorporates the semantic state $\mathbb{S}_{D}^t$, which is formed by previously decoded tokens through the attention mechanism.

We have designed our channel encoder-decoder blocks using a single fully connected (FC) layer. The FC layer in the channel encoder block has an input dimension of $\mathcal{D}_{emb}$ and an output dimension of $2 \mathcal{D}$, whereas it is transposed in the channel decoder block. In both blocks, the FC layer is followed by normalization and a PReLU activation function.

\subsection{Semantic Predict-and-Forward (SPF)}
\label{sec:semantic_predict_and_forward}
SPF  is similar to the SLF described in section \ref{sec:semantic_lossy_forwarding} in terms of the operations done at the source and the destination, but differs in the operation done at the relay. In contrast to equation \ref{eq:slf_relay_decoding}, we utilize $\mathbf{Y}_{S_2}^t$ in conjunction with the semantic state $\mathbb{S}_R^t$ to predict the source token $w^{t+1}$ at the relay as $\hat{w}_{R}^{t+1}$.

\begin{equation}
    \label{eq:spf_relay_decoding}
    \hat{w}_{R}^{t+1} = S_{R}^{-1} \biggl(\Gamma_{R}^{-1} \Bigl(\mathbf{Y}_{S_2}^t \Bigl), \mathbb{S}_{R}^t \biggl)
\end{equation}

To recall, $\Gamma_{R}^{-1}(\mathbf{Y}_{S_2}^t)$ is the noisy embedding of the token $w^t$ and $\mathbb{S}_{R}^t$ is formed by the noisy embeddings received prior. Hence, SPF tries to predict the next token $w^{t+1}$ based on the noisy embeddings of the previous tokens. 

The encoding at the source is performed in the same manner as described in equation \ref{eq:source_process}. However, for the relay semantic decoder to predict the first token $w^1$, embedding of the [CLS] token of the BERT output, denoted as $S^0(\mathbf{s})$, is encoded and exclusively transmitted from source to relay in a point-to-point (P2P) manner at $t=0$. This initial step enables the relay semantic decoder to grasp the context of the message and predict the first token of the source message. The predictions at the relay, $\hat{w}_{R}^t$ for $t \geq 1$, are encoded and sent to the destination the same way as in equation \ref{eq:relay_encoding}. Since there is no token to predict at $t=T$, the BERT output of the last token, $S^T(\mathbf{s})$, is exclusively sent from source to destination in a point-to-point manner. At the destination, decoding is carried out in the same manner as described in equation \ref{eq:destination_decoding} at every index $t$.

\begin{figure}
\centering
\includegraphics[width=\columnwidth]{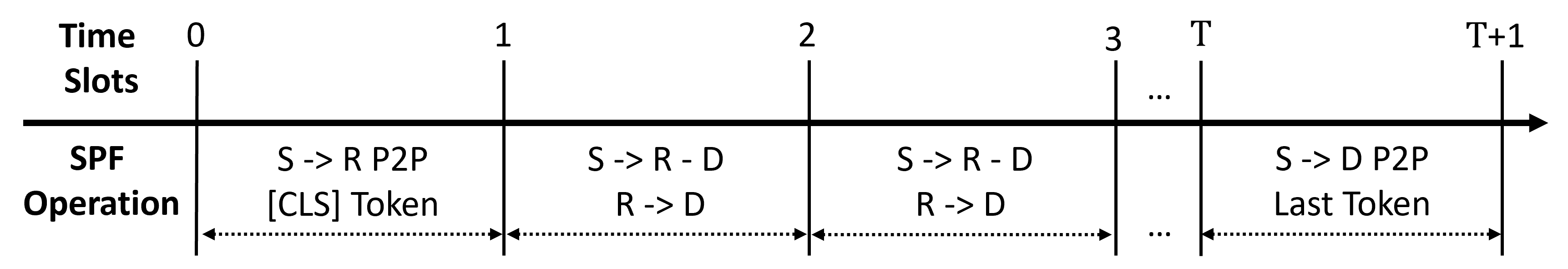}
\caption{Utilization of time slots by SPF.}
\label{fig:time_slots} 
\end{figure}

The communication process is illustrated in figure \ref{fig:time_slots} for a message containing T tokens. The SPF model is based on the orthogonality of Source-Destination and Relay-Destination links. After transmitting the [CLS] token to the relay in P2P, the communication proceeds as a combination of broadcast and an orthogonal multiple access channel (MAC). Specifically, the source broadcasts the signal to the relay and destination, while the relay transmits its prediction to the destination through an orthogonal multiple access channel (MAC). This enables the relay to proactively forward the source message.

\section{Results}
\label{sec:results}
In this section, we evaluate the performances of the SLF and SPF models proposed in sections \ref{sec:semantic_lossy_forwarding} and \ref{sec:semantic_predict_and_forward}. We also compare them against a semantic-agnostic conventional baseline.

\subsection{Simulation Setup}
We perform our simulations over AWGN and Rayleigh fading channels as explained in section \ref{sec:system_model} with parameters in table \ref{table:channel_setting_table}. This setting results in an SNR of $-10\log_{10}({d_{\text{Link}}^{\alpha}}\times{{\sigma_Z}^2})$ dB. We employed transmit power control in the form of channel inversion. 
We have used the European Parliament Corpus dataset, which consists of approximately two million English sentences \cite{europarl}. In our pre-processing, we have excluded sentences containing non-ASCII characters and those with fewer than five tokens. Tokenization is performed using the BERT tokenizer, and sentences exceeding 30 tokens are truncated. The dataset is then split into training (70\%), validation (15\%), and test (15\%) sets. This process has resulted in 24045 unique tokens, which is the vocabulary size in our proposed models.

\renewcommand{\arraystretch}{1.25} 
\begin{table}
\centering
\caption{Simulation Parameters for Channel Models}
\begin{tabular}{|l|c|}
\hline
\textbf{Parameter}           & \textbf{Value} \\ \hline
Noise Power Spectral Density, $N_0$ & -174 dBm/Hz    \\ \hline
Channel Bandwidth, $W$           & 1 MHz           \\ \hline
Transmit Power Constraint, $p_T$      & 30 dBm            \\ \hline
Path Loss Exponent, $\alpha$           & 4                  \\ \hline
\end{tabular}
\label{table:channel_setting_table}
\end{table}
\renewcommand{\arraystretch}{1} 

\renewcommand{\arraystretch}{1.25} 
\begin{table}
\centering
\caption{Training Parameters}
\begin{tabular}{|c|c|}
\hline
\textbf{Parameter} & \textbf{SLF} \& \textbf{SPF} \\ \hline
Attention Heads  & {6}    \\ \hline
Num. Transformer Blocks & {6}  \\ \hline
Number of Epochs & {10}   \\ \hline
Learning Rate & {5e-4}    \\ \hline
Channel Dimension, $2\mathcal{D}$ & {256} \\ \hline
$d_{min}$ &   {2 km}  \\ \hline
$d_{max}$ & {7 km}  \\ \hline
$\gamma_{min}$ & {0.2}  \\ \hline
$\gamma_{max}$ & {0.8}  \\ \hline
Weight Decay & {0.01}  \\ \hline
Embedding Dimension, $\mathcal{D}_{emb}$ &{384}  \\ \hline
Vocab. Size &{24045}  \\ \hline
\end{tabular}
\label{table:training_parameters}
\end{table}
\renewcommand{\arraystretch}{1} 

\begin{algorithm}
\setstretch{1.1}
\caption{Training Semantic Decoder} \label{alg:relay_semantic_decoder}
\begin{algorithmic}
\STATE \textbf{Inputs: } Parameters in table \ref{table:training_parameters}

\SetKwFunction{MyFunction}{Train Semantic Decoder}
  \SetKwProg{Fn}{Function}{:}{}
  \Fn{\MyFunction{}}{
    \For{every batch}{
        \For{$t\gets1$ \KwTo $L$}{    
            \STATE $\mathbf{X}_S^t = S^t \bigl(\mathbf{s} \bigl) $
            \STATE $    \hat{w}_{D}^t = S_{D}^{-1} (\mathbf{X}_S^t, \mathbb{S}_D^t)$
            \STATE $\mathbb{S}_D^{t+1}$ = update\_state($\mathbb{S}_D^t$, $\hat{w}_D^t$)
        }
    \STATE $AdamW(params, CE(\mathbf{s}, \hat{\mathbf{s}}))$
    }
  }
\end{algorithmic}
\end{algorithm}

\begin{algorithm}[t]
\setstretch{1.1}
\caption{Training Source TX - Relay RX} \label{alg:src_relay_training}
\begin{algorithmic}
\STATE \textbf{Inputs:} Parameters in table \ref{table:channel_setting_table}, and \ref{table:training_parameters}

\SetKwFunction{MyFunction}{get\_distance}
  \SetKwProg{Fn}{Function}{:}{}
  \Fn{\MyFunction{$d_{min}$, $d_{max}$}}{
    \KwRet  Uniform RV($d_{min}$, $d_{max}$) \;
  }
  
\SetKwFunction{MyFunction}{Train Source TX - Relay RX}
  
  \SetKwProg{Fn}{Function}{:}{}
  \Fn{\MyFunction{}}{
    \For{every batch}{
        \STATE $d_{SD}$ = get\_distance($d_{min}$, $d_{max}$)
        \STATE $d_{SR}$ = get\_distance($d_{SD} \times \gamma_{min}$, $d_{SD} \times \gamma_{max}$)
        
        \For{$t\gets1$ \KwTo $L$}{    
            \STATE $\mathbf{X}_S^t =  \Gamma_{S} \bigl(S^t\bigl(\mathbf{s} \bigl) \bigl) $
            \STATE $\mathbf{Y}_{S_2}^t = h_{SR}\mathbf{X}_S^t + \mathbf{Z}$
            \STATE $\hat{w}_R^t = S_{R}^{-1} \bigl(\Gamma_{R}^{-1} (\mathbf{Y}_{S_2}^t), \mathbb{S}_R^t \bigl)$
            \STATE $\mathbb{S}_R^{t+1}$ = update\_state($\mathbb{S}_R^t$, $\hat{w}_R^t$)

        }
    \STATE $AdamW(params, CE(\mathbf{s}, \hat{\mathbf{s}}))$
    }
  }
\end{algorithmic}
\end{algorithm}
We 
have trained our proposed models following the procedures outlined in algorithms \ref{alg:relay_semantic_decoder}, \ref{alg:src_relay_training}, and \ref{alg:relay_dst_training}. Initially, semantic decoders at the relay and the destination are trained in a noiseless environment using algorithm \ref{alg:relay_semantic_decoder}. Source TX - relay RX block is then trained in a noisy environment employing algorithm \ref{alg:src_relay_training}. Finally, the trained source TX - relay RX block is frozen, and relay TX - destination RX block is trained in a noisy environment using algorithm \ref{alg:relay_dst_training}. All algorithms utilize the AdamW optimization algorithm, and detailed parameters are provided in table \ref{table:training_parameters}. The cross-entropy loss function, denoted as $CE(.,.)$, is used for all algorithms. Notably, algorithms \ref{alg:relay_semantic_decoder} and \ref{alg:src_relay_training} differ in the training of SLF and SPF. For SLF, the input of $S_D^{-1}$ and $S_R^{-1}$ depends on the time $t$, while for SPF, it depends on $t-1$ to obtain $\hat{w}_D^t$ and $\hat{w}_R^t$, respectively. During the training of both SLF and SPF, we incorporate a one-cycle learning rate scheduler to enhance convergence.

\begin{algorithm}
\setstretch{1.1}
\caption{Training Relay TX - Destination RX} \label{alg:relay_dst_training}
\begin{algorithmic}
\STATE \textbf{Inputs:} Parameters in table \ref{table:channel_setting_table}, and \ref{table:training_parameters}
  
\SetKwFunction{MyFunction}{Train Relay TX - Destination RX}
  
  \SetKwProg{Fn}{Function}{:}{}
  \Fn{\MyFunction{}}{
    \For{every batch}{
        \STATE $d_{SD}$ = get\_distance($d_{min}$, $d_{max}$)
        \STATE $d_{SR}$ = get\_distance($d_{SD} \times \gamma_{min}$, $d_{SD} \times \gamma_{max}$)
        \STATE $d_{RD}$ = $d_{SD}$ - $d_{SR}$ 
        
        \For{$t\gets1$ \KwTo $L$}{    
            \STATE $\mathbf{X}_S^t =  \Gamma_{S} \Bigl(S^t \bigl(\mathbf{s} \bigl) \Bigl) $
            \STATE $    \mathbf{Y}_{S_2}^t = h_{SR}\mathbf{X}_S^t + \mathbf{Z}$
            \STATE $\mathbf{X}_R^t$ = $Relay\Bigl(\mathbf{Y}_{S_2}^t\Bigl)$
            \STATE $    \mathbf{Y}_{S_1}^t = h_{SD}\mathbf{X}_{S}^t + \mathbf{Z}$
            \STATE $    \mathbf{Y}_R^t = h_{RD}\mathbf{X}_R^t + \mathbf{Z}$
            \STATE $    \hat{w}^t = S_{D}^{-1}(\Gamma_{D}^{-1} (\mathbf{Y}_{S_1}^t, \mathbf{Y}_R^t), \mathbb{S}_{D}^t)$
            \STATE $\mathbb{S}_D^{t+1}$ = update\_state($\mathbb{S}_D^t$, $\hat{w}^t$)
        }
    \STATE $AdamW(params, CE(\mathbf{s}, \hat{\mathbf{s}}))$
    }
  }
\end{algorithmic}
\end{algorithm}

\begin{figure*}
    \centering
  \subfloat[BLEU 3-gram vs. Relay position for an Source-Destination distance of 4000m over AWGN channel. \label{bleu3_sweep_relay_AWGN}]{\includegraphics[width=0.355\linewidth]{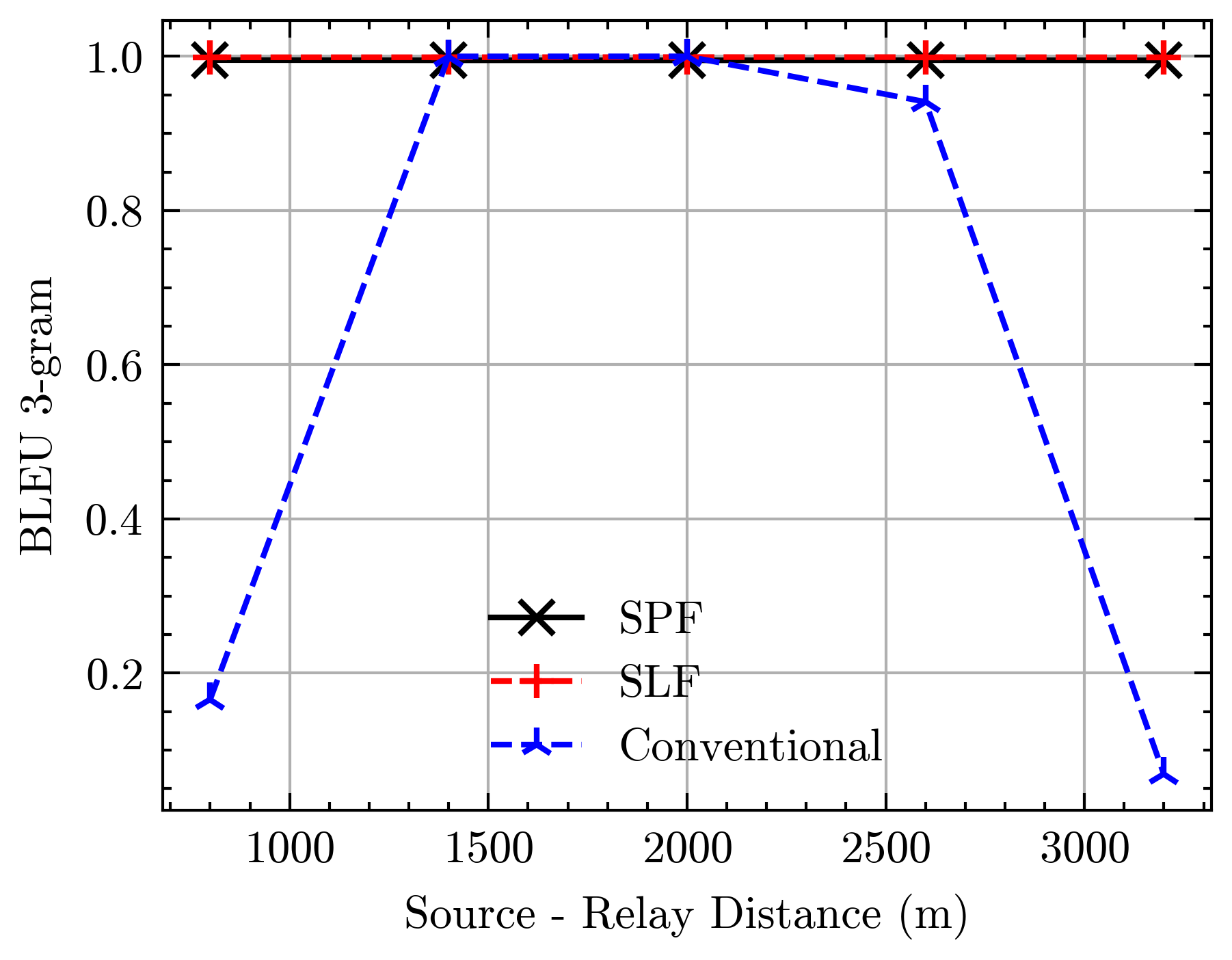}}
    \hspace{10mm}
  \subfloat[BLEU 3-gram vs. Relay position for an Source-Destination distance of 4000m over Rayleigh fading channel. \label{bleu3_sweep_relay_Rayleigh}]{\includegraphics[width=0.355\linewidth]{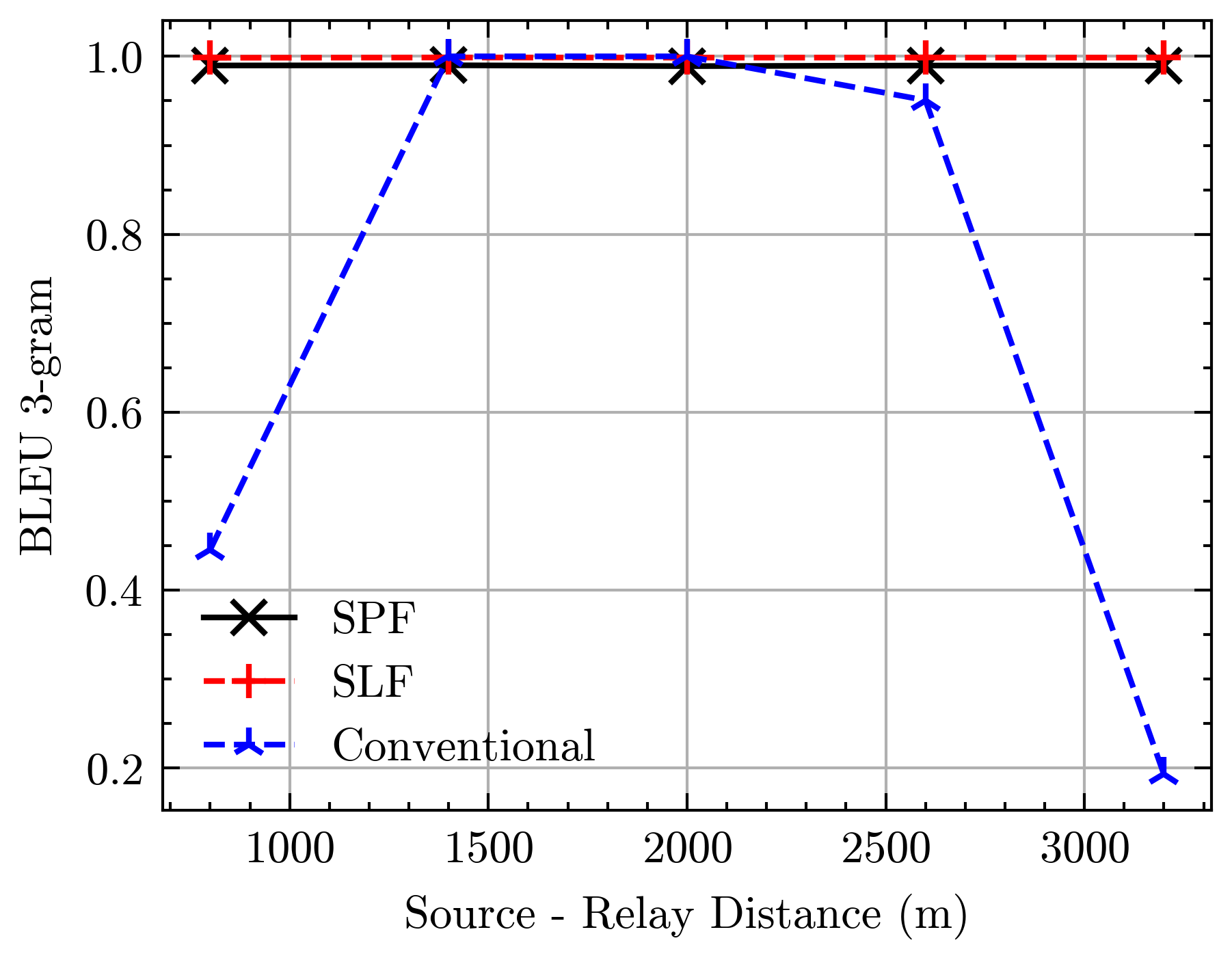}}
  \\
  \subfloat[Semantic Similarity vs. Relay position for an Source-Destination distance of 4000m over AWGN channel.\label{semsim_sweep_relay_AWGN}]{\includegraphics[width=0.355\linewidth]{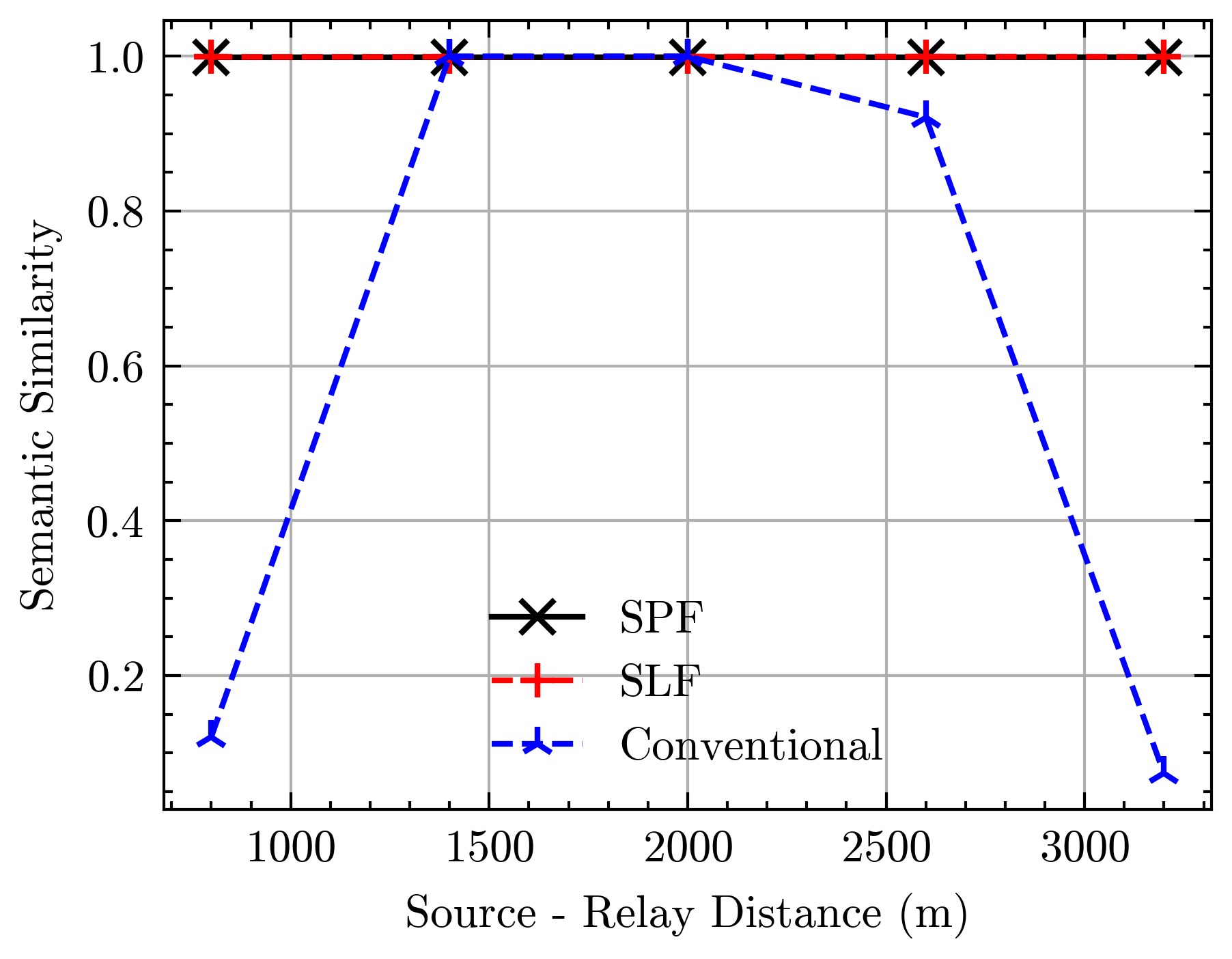}}
      \hspace{10mm}
  \subfloat[Semantic Similarity vs. Relay position for an Source-Destination distance of 4000m over Rayleigh fading channel.\label{semsim_sweep_relay_Rayleigh}]{\includegraphics[width=0.355\linewidth]{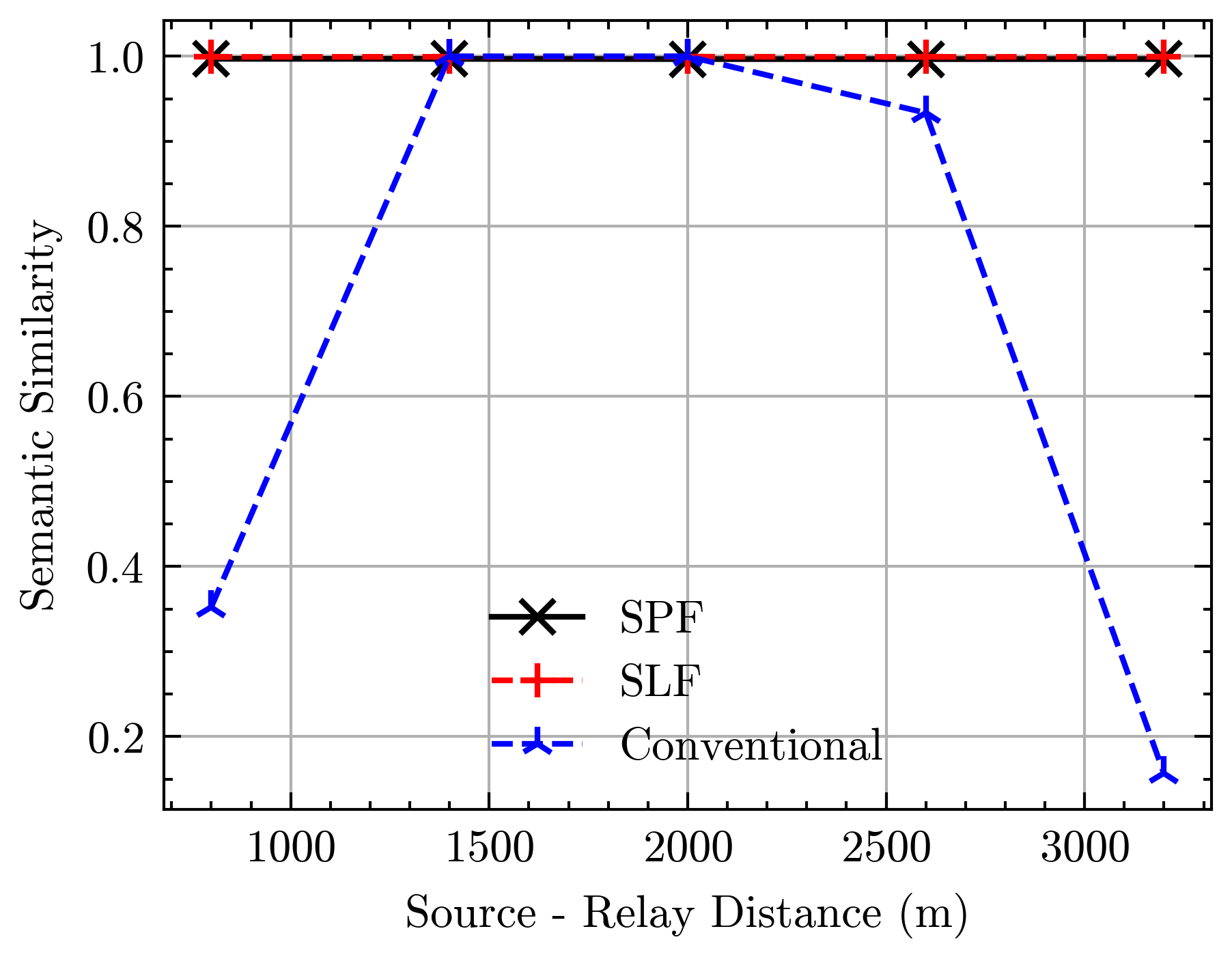}}
  \caption{Performance results for different relay positions.}
  \label{fig:relay_sweep_results} 
\end{figure*}

\begin{figure*}
    \centering
  \subfloat[BLEU 3-gram vs. Source-Destination distances with relay node in the middle over AWGN channel. \label{bleu3_sweep_sd_AWGN}]{\includegraphics[width=0.328\linewidth]{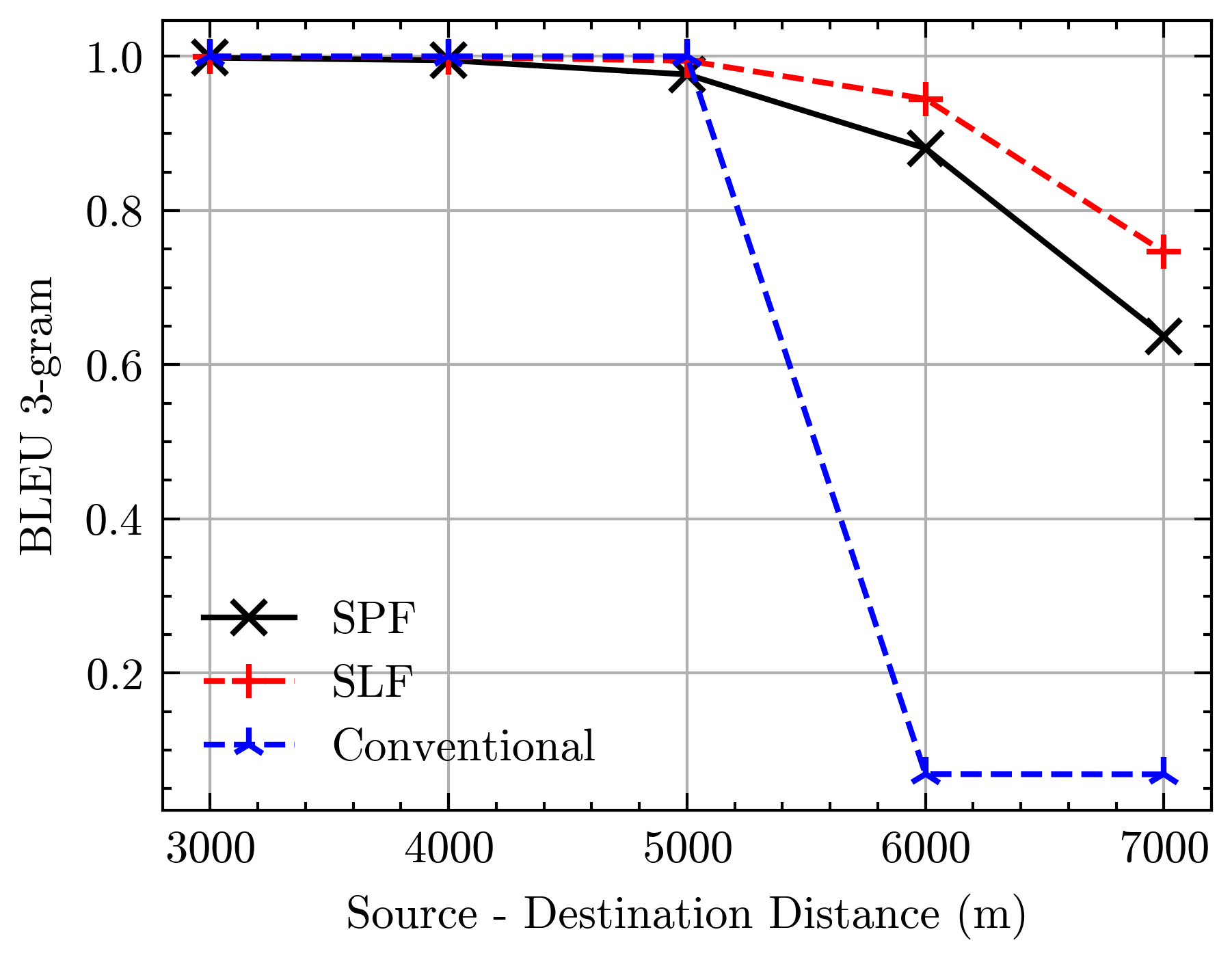}}
    \hspace{10mm}
      \subfloat[BLEU 3-gram vs. Source-Destination distances with relay node in the middle over Rayleigh fading channel. \label{bleu3_sweep_sd_Rayleigh}]{\includegraphics[width=0.328\linewidth]{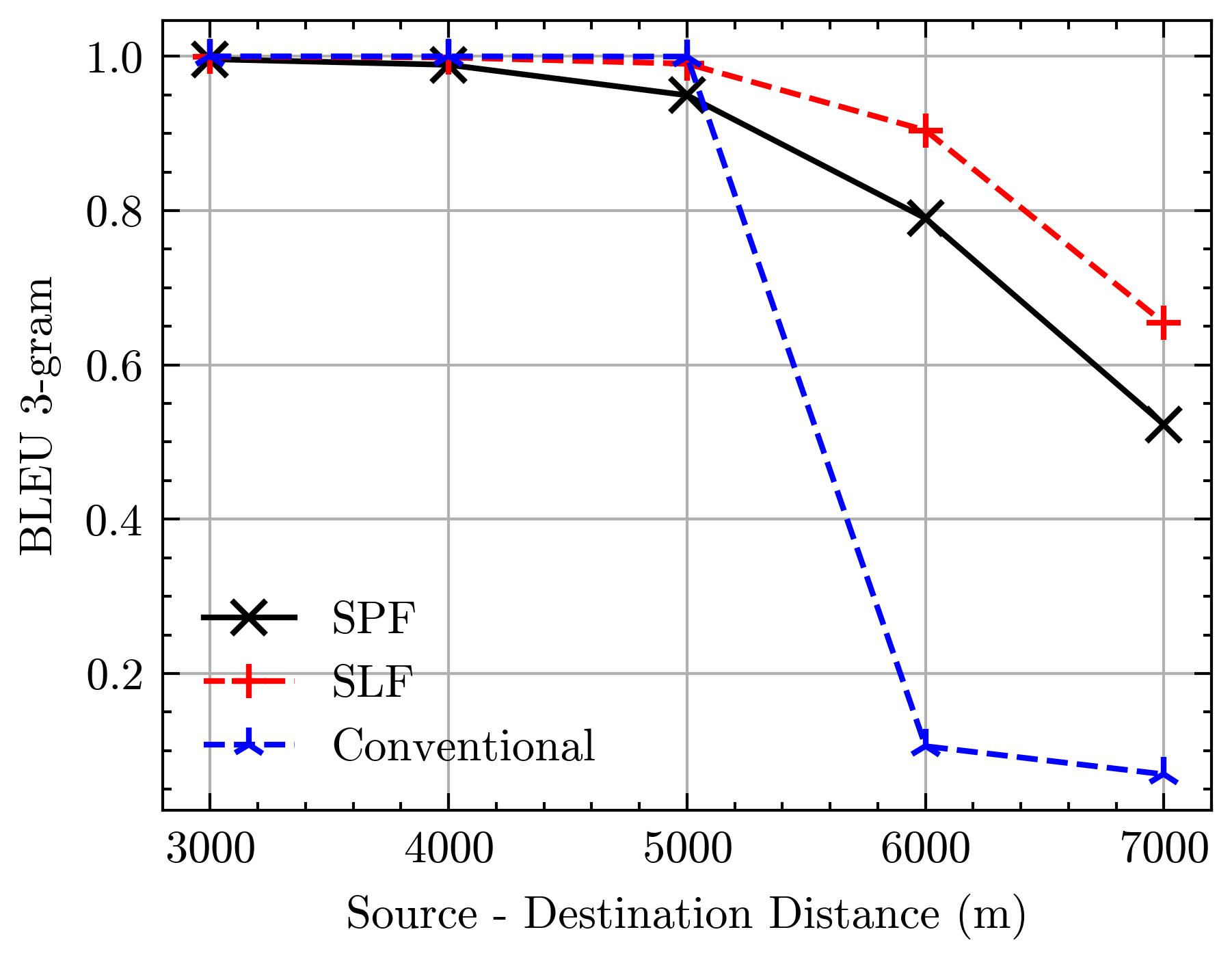}}
  \\
  \subfloat[Semantic Similarity vs. Source-Destination distances with relay placed in the middle over AWGN channel. \label{semsim_sweep_sd_AWGN}]{\includegraphics[width=0.328\linewidth]{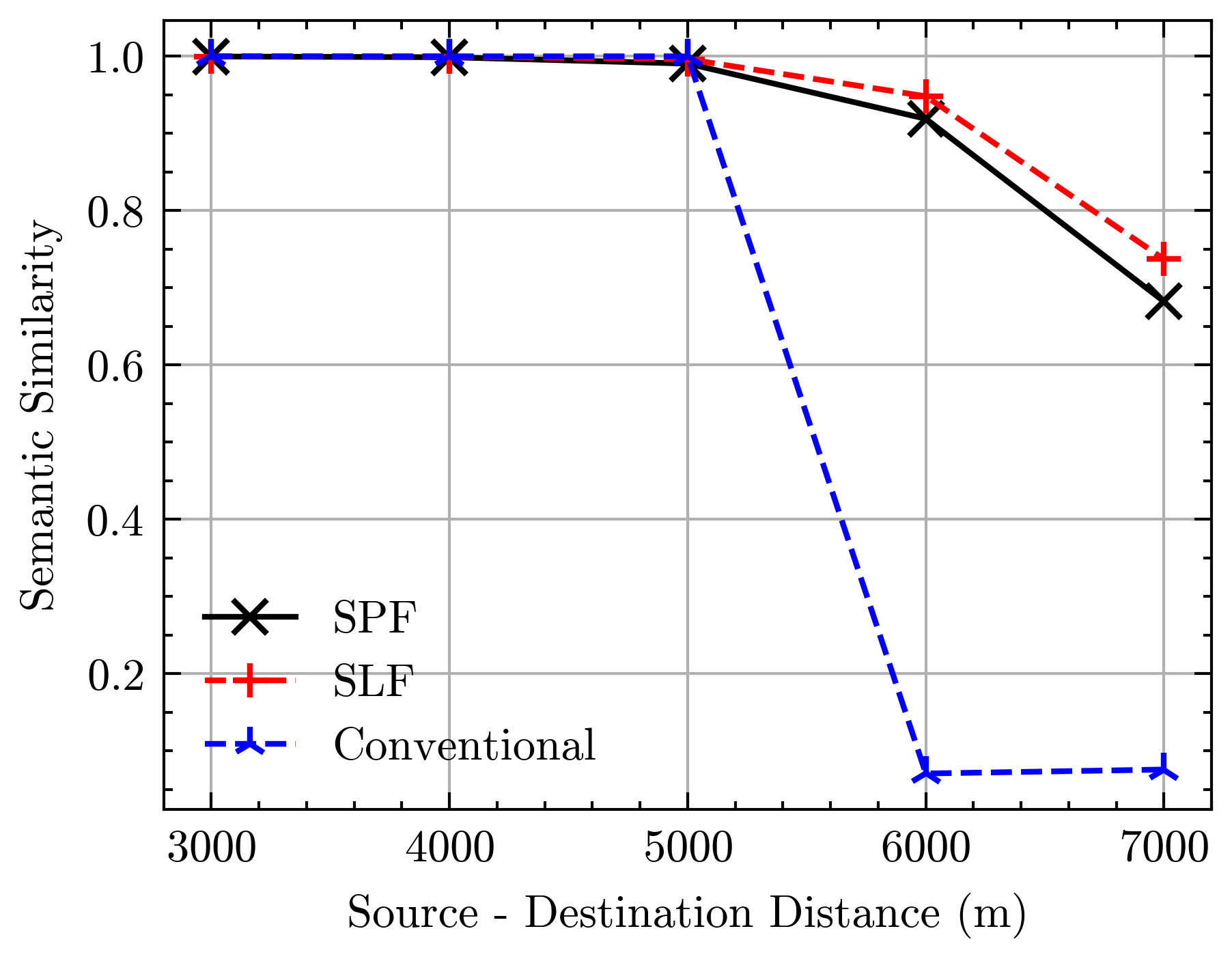}}
      \hspace{10mm}
  \subfloat[Semantic Similarity vs. Source-Destination distances with relay placed in the middle over Rayleigh fading channel. \label{semsim_sweep_sd_Rayleigh}]{\includegraphics[width=0.328\linewidth]{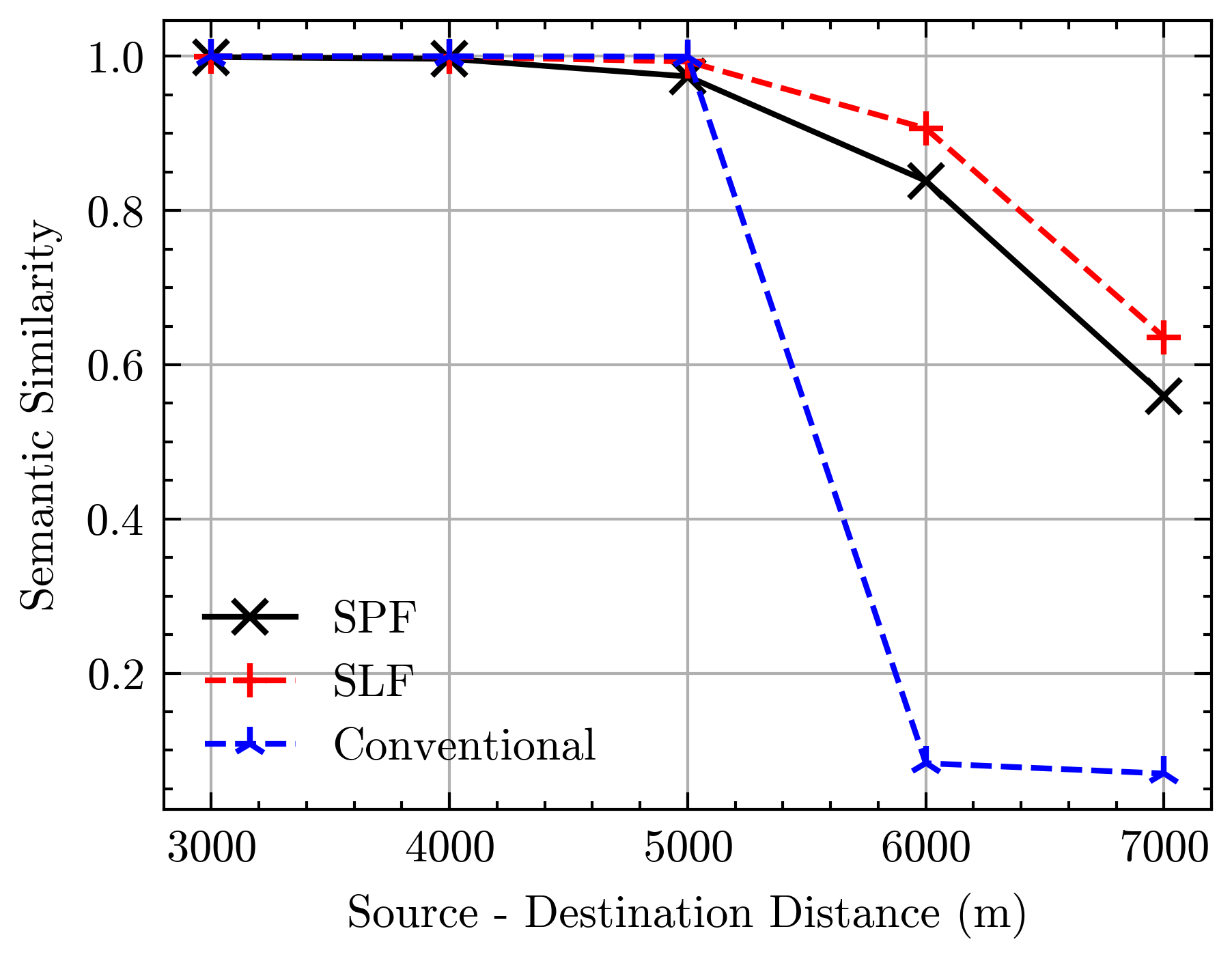}}
  \caption{Performance results for different Source-Destination distances.}
  \label{fig:src_dst_sweep_results} 
\end{figure*}

As the baseline model, we have employed a conventional decode-and-forward (DF) relay with fixed-length source coding, Reed-Solomon (RS) channel coding, and QPSK modulation. The decoding at relay and destination is done based on the maximum likelihood (ML) rule where all the channel state information is assumed to be known. The ML rule at the destination is performed as follows.

\begin{equation}
\begin{split}    
        \label{eq:maximum_likelihood_detection}
        \hat{\mathbf{x}}_S 
        &= \argmax_{\mathbf{x}_S \in \mathcal{X}} \text{Pr}(\mathbf{y}_{SD} \mid \mathbf{x}_{S}) \text{Pr}(\mathbf{y}_{RD} \mid \mathbf{x}_{S}) \\
        &= \argmax_{\mathbf{x}_S \in \mathcal{X}} \text{Pr}(\mathbf{y}_{SD} \mid \mathbf{x}_{S}) \sum_{\mathbf{x}_R \in \mathcal{X}} \text{Pr}(\mathbf{x}_R \mid \mathbf{x}_S) \text{Pr}(\mathbf{y}_{RD} \mid \mathbf{x}_R) \\
\end{split}
\end{equation}

\noindent Where $\mathbf{x}_R, \mathbf{x}_S \in \mathcal{X} = \{(-1, -1), (-1, 1), (1, -1), (1, 1)\}$ and $\mathbf{y}_{RD}$, $\mathbf{y}_{SD}$ are the received signals from the relay and the source, respectively. The transition probabilities $\text{Pr}(\mathbf{x}_R \mid \mathbf{x}_S)$ are calculated empirically on the same training dataset. 

To have a fair comparison, the channel coding rate is selected to keep the number of channel uses per token the same. In our proposed models $2\mathcal{D}$ dimensional vector is sent in $\mathcal{D}$ channel uses using I/Q modulation, where $\mathcal{D}=128$ as given in table \ref{table:training_parameters}. To attain the same rate in the conventional baseline with QPSK modulation, 256 bits can be spared for each token. Since 15 bits can represent 24045 unique tokens with fixed-length source coding, we employ RS(255, 15) channel coding.

\subsection{Performance Metric}
\label{sec:performance_metric}
We evaluate our models using semantic similarity and the BLEU metric \cite{zhijin_deepsc, bleu_score}. The BLEU score enables a comparison of the wording between two sentences, offering insights into the fidelity of message reconstruction. However, due to its lack of intuition regarding meaning conveyance, we complement this metric with semantic similarity. To assess semantic similarity, we employ the SBERT model, which provides semantic embeddings for sentences \cite{sentence-bert}. The similarity score is calculated by the cosine similarity between the SBERT embeddings of given sentences, resulting in a score between $[-1, 1]$, where 1 indicates that two sentences have exactly the same meaning.

\subsection{Numerical Results}
We simulate our system performance varying the relay's location. In figure \ref{fig:relay_sweep_results}, we observe that the performance of SLF and SPF models remains unchanged. We attribute this invariance to the machine-learning approach we have taken. Since the relay position is randomly chosen for every batch during our training, models learn to generalize to unknown relay positions. This is more evident when we compare it with the conventional baseline, which attains its optimal performance when the relay is positioned in the middle, i.e., $d_{SR} = d_{RD} = 0.5d_{SD}$. 

In figure \ref{fig:src_dst_sweep_results}, we observe that proposed models significantly outperform the conventional model. Notably, the SPF model achieves performance comparable to the SLF. This observation underscores SPF's ability to accurately predict the next tokens. Even in cases of large source to destination distances, proactive relaying through prediction does not lead to a significant degradation in performance. 

\section{Conclusion}
\label{sec:conclusion}
In this paper, we have investigated cooperative text communications involving a relay node. We have proposed two relaying techniques that utilize an attention mechanism to establish a dynamic semantic state at the relay node. In SLF, this state is employed to decode and re-encode the current signal. In SPF, we use this state to perform predictions to enable proactive forwarding. Our simulation results demonstrate that both models outperform the conventional semantic-agnostic baseline. Additionally, we have observed that the proposed prediction scheme can accurately predict the next source signal, warranting further investigation. For example, proactive forwarding can help the destination to decode more than one token per embedding or to decode the entire sentence from a single embedding.

Semantic cooperative communications is yet at its nascence and interesting future directions remain including the role of multiple cooperating relays in extracting semantics, and task-oriented cooperative communications assisted by semantic communications.

\bibliographystyle{IEEEtran}{}
\bibliography{semantic-comm-refs}
\end{document}